\documentclass{optica-article}
\journal{opticajournal}
\articletype{Research Article}

\usepackage{physics}
\usepackage{glossaries}
\usepackage{siunitx}
\usepackage{bm}

\newacronym{cem}{CEM}{computational electromagnetics}
\newacronym{cmt}{CMT}{coupled-mode theory}
\newacronym{dcf}{DCF}{double-clad fiber}
\newacronym{fdtd}{FDTD}{finite-difference time-domain}
\newacronym{itr}{ITR}{inverse taper ratio}
\newacronym{lp}{LP}{linearly polarized}
\newacronym{mom}{MoM}{the method of moments}
\newacronym{mspl}{MSPL}{mode-selective photonic lantern}
\newacronym{na}{NA}{numerical aperture}
\newacronym{os}{OS}{optical switch}
\newacronym{pl}{PL}{photonic lantern}

\begin{document}

\title{Three-Mode Photonic Lanterns: Comprehensive Analysis from Theory to Experiments}

\author{Rodrigo Itzamn\'a Becerra-Deana,\authormark{1,2} Raphael Maltais-Tariant,\authormark{1,3} Guillaume Ramadier,\authormark{1} Martin Poinsinet de Sivry-Houle,\authormark{1} St\'ephane Virally,\authormark{1} Caroline Boudoux,\authormark{1,2,3} and Nicolas Godbout\authormark{1,2,*}}

\address{\authormark{1}Polytechnique Montr\'eal, 2500 Chemin de Polytechnique, Montr\'eal, QC H3T 1J4, Canada\\
\authormark{2}Castor Optics, 361 Boulevard Montpellier, Saint-Laurent, QC H4N 2G6, Canada\\
\authormark{3}CHU Sainte-Justine, 3175 Chemin de la C{\^o}te Sainte-Catherine, Montr{\'e}al, QC H3T 1C5, Canada}

\email{\authormark{*}nicolas.godbout@polymtl.ca}

\begin{abstract*}
    The design space for photonic lanterns is large and complex, making it challenging to identify optimal parameters to achieve specific performances, such as coupling, bandwidth, and insertion loss.
    Effectively navigating this space requires modeling tools capable to extract the most characterizing parameters.
    This work contrasts theoretical modeling with experimental realizations of the four possible types of $3\times1$ photonic lanterns using double-clad fibers, covering a spectrum from conventional to hybrid to mode-specific configurations.
    This work highlights the experimental characteristics of each photonic lantern.
\end{abstract*}

\section{Introduction}\label{sec:Introduction}
    \Glspl{pl} are $N\times1$ transverse-mode division multiplexers/de-multiplexers composed of $N$ single-mode fibers fused together into a few-mode waveguide~\cite{birks_photonic_2015}.
    They are typically fabricated by heating and tapering a bundle of $N$ fibers before cleaving the resulting fused section, acting as a few-mode segment.
    \Glspl{pl} are used in a variety of applications including telecommunications~\cite{tele2QIU2023109229,teleLEONSAVAL201746,tele3JIANG2020125988,Comu4,Comu5, Charactericomu,Comunine,Comuten}, lasers~\cite{LasersLIU2025111501,Lasers2ZHANG2024111180}, astrophysics~\cite{olaya_161_2012,Astro3}, sensing~\cite{FocalplaneLin:23,SensingGU2024131118}, orbital angular momentum mode generation~\cite{OrbitalLu_2019},  efficient collection of light~\cite{ozdur_free-space_2013},  and biomedical imaging~\cite{BIoSalit:20,Maltais-Tariant:23,sivry-houle_all-fiber_2021,Biofive}.

    The characteristics of \glspl{pl} vary greatly depending on the fiber configuration and type within the bundle.
    In general, their behavior can be sorted into three main categories, which we herein call conventional, mode-selective, and hybrid.
    Conventional \glspl{pl} are fabricated from $N$ identical single-mode fibers~\cite{leon-saval_multimode_2005,nonmodeselec,Xu:24} and equally couple each fiber's fundamental modes to the transverse modes of the multimode section.
    Conversely, they enable the equipartite collection of the amplitudes of the transverse modes into all $N$ single-mode outputs without distinction between each port.
    This behavior differs from that of so-called \glsreset{mspl}\glspl{mspl}, consisting of $N$ different fibers acting as mode sorters whereby each transverse mode of the few-mode section is selectively coupled to a specific single-mode fiber~\cite{shen_highly_2018, sivry-houle_all-fiber_2021,EMESUNDER2020102219,Yu:16}.
    \Glspl{mspl} usually have high modal isolation, and their operation is typically wavelength-independent, up to a bandwidth of $500~\unit{nm}$~\cite{fontaine_photonic_2022}.
    The hybrid type---also called semi-selective or group-mode-selective~\cite{leon-saval_mode-selective_2014, velazquez-benitez_scaling_2018,GMSPL1}---refers to \glspl{pl} behaving in between the conventional and mode-selective types.

    The plurality of \gls{pl} designs provides a large design space.
    \Gls{cem} methods~\cite{methods1, Methods2Sharma:89} such as \gls{fdtd} and \gls{mom} are often used when no simple closed-form solutions to Maxwell's equations exist. Recent studies have demonstrated that \Gls{cmt} serves as an effective implementation for simulating guided waves in complex structures~\cite{bures_guided_2008, Chen:20, CMT3, deSivry-Houle:24}. This method aptly captures the complexities of electromagnetic field dynamics, transforming it into a more manageable eigenvalue problem. Our group accomplished this through a Python package named SuPyMode~\cite{deSivry-Houle:24}. 
    
    \Gls{cmt} modeling of \glspl{pl} allows observing the spectrum of behaviors exhibited by various types of \glspl{pl}, particularly those stemming from symmetry-breaking fiber arrangements, which is well represented by eigenmode decomposition. However, most of the existing simulators operate under ideal conditions. While they provide valuable approximations, fully understanding a device’s performance requires actual fabrication and characterization to ascertain the modal coupling and wavelength response, which are crucial for implementation in various fields.
    
    In this study, we first briefly introduce adiabatic criteria and the logarithmic taper slope associated with a particular fabrication sequence.
    We combine these two concepts to determine the best parameters for a given component. We then compare theoretical predictions with actual fabricated components of each type of photonic lantern, demonstrating their different performances at each port over a wide bandwidth and their optical profiles. We emphasize the significant differences in the experimental behaviors of components that may seem very similar at first but have distinct internal symmetries.

\section{Modeling tools}\label{sec:Methods}
    The main characteristics of \glspl{pl} can be determined theoretically by comparing the inter-modal adiabatic criteria to the logarithmic slope of the taper.
    We briefly introduce those two concepts and their relation to the final characteristics of the devices.

\subsection{Adiabatic criterion}\label{sec:criterion}
    The design of \glspl{pl} involves many parameters, including the types of fibers, stacking arrangement, amount of initial fusion, and tapering profile.
    All of these parameters directly influence the adiabatic criteria, which ultimately determine if and how a device can be fabricated.

\begin{figure}[t]
    \centering
    \includegraphics[width=\textwidth]{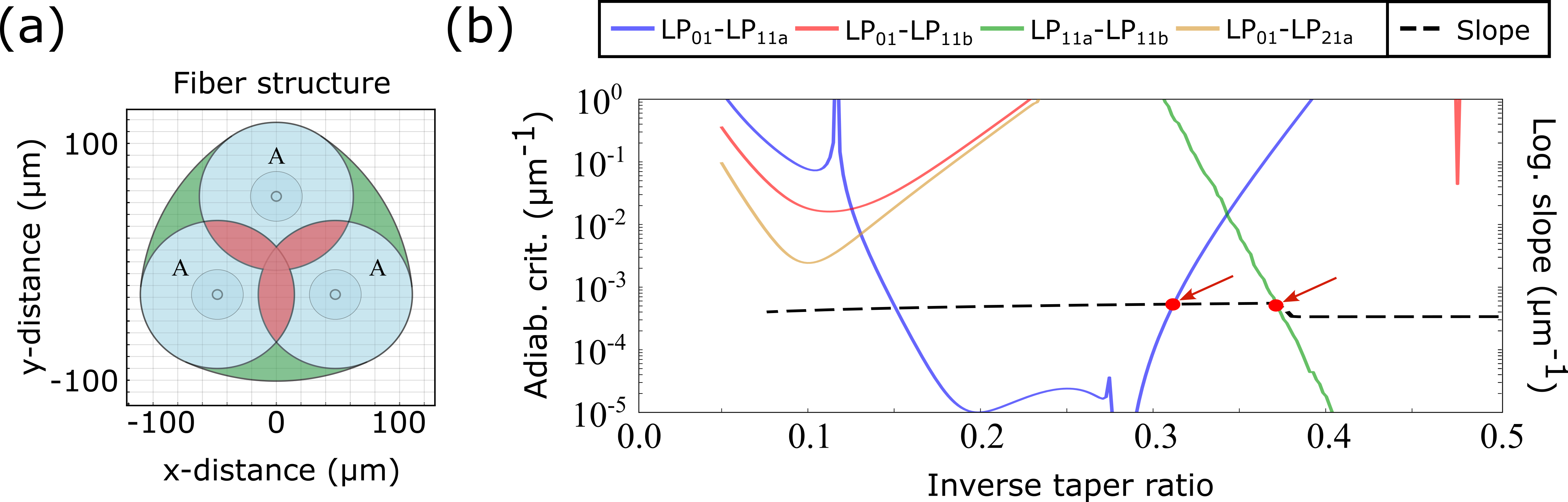}
    \caption{Simulation of a three-mode structure using SuPyMode. (a) Cross-sectional map of refraction indices after fusion initial fusion of the three-fiber bundle.
    (b) Adiabatic criteria for each mode pair (solid lines) and logarithmic slope of the taper (dashed line) for a given fabrication sequence.
    The arrows indicate crossings that potentially imply coupling between specific modes, although additional judgment must be applied (see text and supplementary material-Supplement 1).}
    \label{fig:adiabatic_graphs}
\end{figure}

    In \gls{cmt}, the adiabatic criterion $\overline{\alpha}_{ij}$ between two transverse eigenmodes $i$ and $j$ is a quantity with units of the inverse of a length.
    It depends only on two properties of the modes in question: their propagation constants $\beta_{i}$ and $\beta_{j}$, and their normalized amplitude profiles $\Psi_{i}$ and $\Psi_{j}$~\cite{bures_guided_2008,Chen:20,deSivry-Houle:24}.
    At any position $z$ along the taper, these can be calculated from the transverse refraction index map $n_0(x,y,z)$ of the fused fiber bundle.
    The package that we used in this study, SuPyMode~\cite{martin_poinsinet_de_sivry_houle_2023_10215492,deSivry-Houle:24}, determines this map from homotetic transformation of the cross-sectional map $n_0(x,y)$ of the fused fiber bundle before tapering. A typical map is shown in Fig.~\ref{fig:adiabatic_graphs}(a).
    The homotetic ratio $\bm{I}\!(z)$ between the length of any line segment on the fiber cross-section before and after tapering at any particular position $z$ is called the \gls{itr}.

\subsection{Logarithmic slope}\label{sec:slope}
    By definition, the logarithmic slope of the taper at any position $z$ is the logarithmic derivative $\frac{1}{\bm{I}}\pdv{\bm{I}}$ of the \gls{itr} at that position. his logarithmic slope, also known as the taper shape, is controlled by the scanning and pulling motors during the fabrication process. The specific shape of the taper is determined by theoretical considerations~\cite{birks_shape_1992}.
    Each step of the sequence consists of heating a section of the bundle, at each point increasing or decreasing the length of the heated section to the value $L=L_0+\alpha\,\delta z$, where $L_0$ is the original heating length of the section, and $\delta z$ is the current elongation of the section.
    The coefficient $\alpha$ is unitless and ranges from -1 to +1.
    A positive value of $\alpha$ indicates that the length of the heated section increases as the step progresses.
    In turn, a negative value of $\alpha$ indicates that the length of the heated section decreases throughout the step.
    The \gls{itr} is fully determined by the ensemble of parameters $L_0$ and $\alpha$~\cite{birks_shape_1992} for each step of the sequence.

\subsection{Coupling threshold}\label{sec:threshold}
    The threshold between inter-modal coupling and inter-modal isolation is determined for each pair of modes $\qty(i,j)$ by comparing the adiabatic criterion for the pair to the logarithmic slope determined by the fabrication sequence.
    Specifically, the condition for inter-modal isolation is $\frac{1}{\bm{I}}\pdv{\bm{I}}{z}\le\overline{\alpha}_{ij}$ over the full length of the taper.

    In addition, a taper is called adiabatic when inter-modal isolation is maintained for all pairs of modes guided by the structure, or
    \begin{equation}
        \frac{1}{\bm{I}\!(z)}\pdv{\bm{I}\!(z)}{z}\le\min_{\qty(i,j)}\overline{\alpha}_{ij}(z)\quad\forall z.
        \label{eq:adiabaticcriterion}
    \end{equation}

    Thus, one of the primordial modeling tools for the design of \glspl{pl} is a graph comparing all relevant adiabatic criteria to the logarithmic slope determined by the fabrication sequence.
    An example of such a graph is shown in Fig.~\ref{fig:adiabatic_graphs}(b).
    This figure shows adiabatic criteria and logarithmic slope values for a specific fabrication sequence as a function of \gls{itr} rather than the longitudinal position $z$ of the taper.
    This improves the readability of the graph, as tapers often feature sections with constant \glspl{itr}.
    The features shown in this particular graph will be reproduced and explained in more details below as it pertains to a case studied in this paper.
    The supplementary material (Supplement 1) provides a deeper dive into these types of graphs and all the information that can be extracted from them.

\section{Methodology}\label{sec:Methodology}
    In the next section, we highlight the three main fiber configurations and fabrication and characterization of $3\times1$ components as they already exhibit the full variety of \glspl{pl}, from conventional to mode-specific and anything in between.

\subsection{Fiber configurations}\label{sec:config}
    We compare three main types of \glspl{pl}: devices made of three identical fibers should result in conventional \glspl{pl}; devices made of three different fibers should result in \glspl{mspl}; devices made of two different types of fibers should be hybrids. 
 \begin{figure}[t]
    \centering
    \includegraphics[width=0.6\columnwidth]{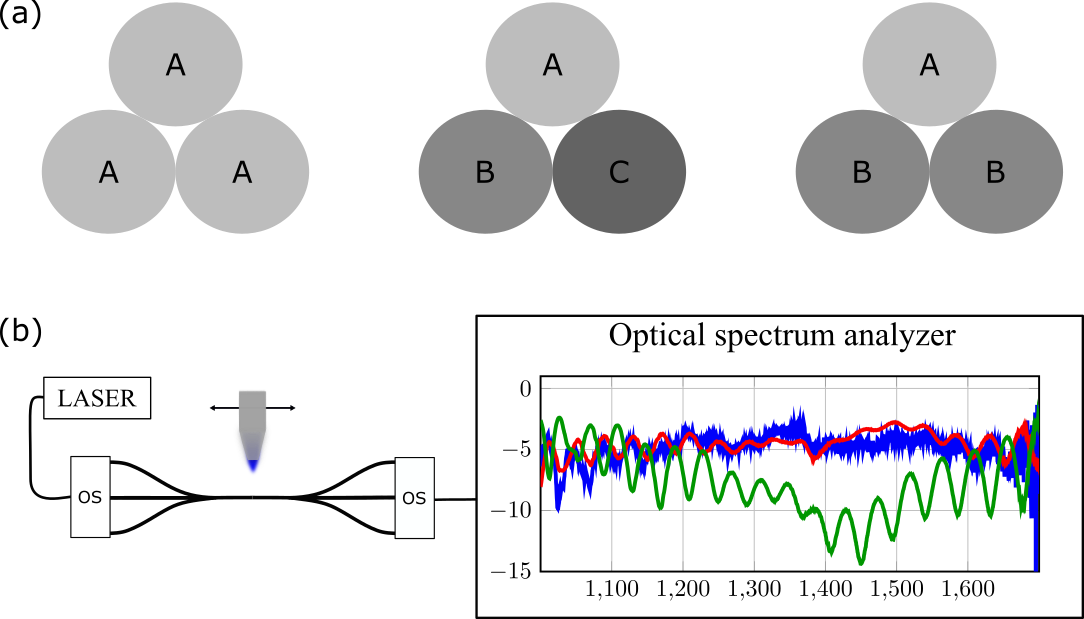}
    \caption{Three-mode structures and fabrication process. 
    (a) Three different structures in which each letter represents a type of fiber; AAA uses three identical fibers, ABC uses different fibers, and ABB uses two types of fibers. 
    (b) Schematic diagram of the fusion and tapering station with a heating torch, an illuminating laser, two optical switches (OS), and an optical spectrum analyzer.}
    \label{fig2}
\end{figure}

    Fig.~\ref{fig2}(a) shows the three main fiber configurations: AAA (three identical fibers), ABC (all different fibers), and ABB (two different types of fibers, each supporting a different mode group).
    For practicality, we use double-clad fibers with single-mode compatibility at 1550~\unit{nm}.
    All fibers feature a 9~\unit{\um} single-mode core and similar \glspl{na}~$\sim0.13$.
    They also feature similar first cladding \glspl{na}~$\sim0.12$ and vary only by first cladding diameters, with respective sizes of 19.6, 32.3, and 42~\unit{\um} (custom-designed, 2058I1, 2058K1, and 2058J1, Universit{\'e} Laval, Prof. Messaddeq, QC, CA).
    
\subsection{Fabrication and characterization}\label{sec:fab}
    Before the fusion-taper procedure, all fiber bundles were constrained inside a synthetic fused silica capillary tube (CV1012, Vitrocom, NJ, USA).
    In-situ, real-time monitoring of the components during fusion and tapering allows to observe and potentially remedy any anomalous loss or coupling between modes~\cite{becerra-deana_mode-selective_2024}.
    The bundles were clamped to an in-house built setup that allows for fusion and tapering using a mobile torch while pulling on each side.
    For better comparison, all devices for this study were fabricated using the same sequence, i.e. the same $\alpha$ and $L_0$ values at all \glspl{itr}.
    In addition, the same characterization setup was used every time.
    Fig.~\ref{fig2}(b) shows the in-situ optical characterization system.
    It starts with a broadband laser source (Energetiq Fiber Coupled Laser-Driven Light Source, model EQ-99, MA, USA) that injects light into a first, in-house built, \gls{os} spliced to the input single-mode fibers of the \glspl{pl}.
    At the output, a second \gls{os} allows the monitoring of each port through an optical spectrum analyzer (OSA, model AQ6317, Ando, Japan). Power transfer is measured by injecting light into one input port and measuring the output across three ports. Additionally, the excess loss is calculated by summing the ratios of all power outputs.
    After fabrication, the few-mode tapered section was cleaved in the center, producing two identical \glspl{pl}.
    Optical far-field profiles were obtained by illuminating each port sequentially with a source operating at 1300~\unit{nm} (HP 8153ASM, Hewlett-Packard, CA, USA) and capturing the image with an infrared camera (SU320KTS-1.7RT, Goodrich, NJ, USA).
   
\section{Results}\label{sec:Results}

The following subsections outline the key characteristics of each type of 3-mode \gls{pl}. This includes a theoretical analysis of the fabrication process, followed by experimental results concerning the final structure's power transfer and the component's optical profile after it has been cleaved. This procedure is consistent across all components, allowing us to identify their main differences.

\begin{figure}[t]
    \centering
    \includegraphics[width=\columnwidth]{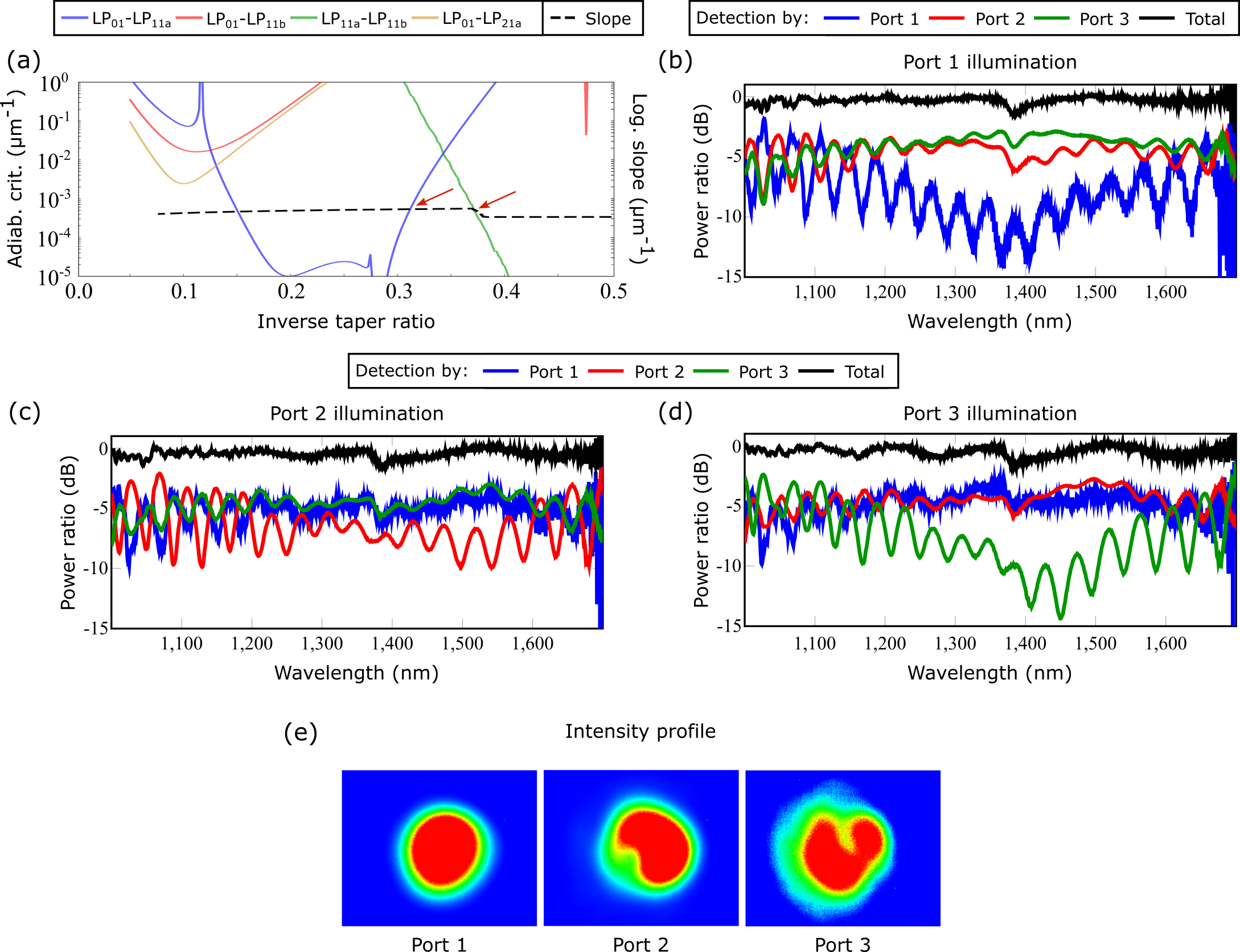}
    \caption{AAA fiber configuration.
    (a) Adiabatic criteria for three identical (AAA) double-clad fibers with a 42~\unit{\um}-diameter first cladding.
    The solid blue, red, and green lines show the criteria for each pair of guided modes (LP$_{01}$, LP$_{11\mathrm{a}}$ and LP$_{11\mathrm{b}}$, respectively) in the few-mode tapered section of the device.
    The lowest adiabatic criterion for coupling between a guided and a non-guided mode (LP$_{01}$ and LP$_{21\mathrm{a}}$ in this instance) is also shown.
    The dashed black curve corresponds to the logarithmic slope for the fabrication sequence.
    (b--d) Power ratio in each branch, with respect to the input power, after fusion and tapering but before cleaving (branch 1 in blue, 2 in red, 3 in green, total in black) when illuminating input 1 (b), input 2 (c), or input 3 (d), respectively.
    (e) Far-field optical profile of the few-mode structure after cleaving and illuminating each port.}
    \label{fig:AAA}
\end{figure}

\subsection{Conventional photonic lanterns}\label{sec:AAA}
    Fig.~\ref{fig:AAA} shows the results for a type AAA component, conventional \glspl{pl}, using three identical fibers (first cladding diameter of 42~\unit{\um}).
    Fig.~\ref{fig:AAA}(a) shows the adiabatic criteria (solid lines) and logarithmic slope (dashed line) calculated by SuPyMode.
    Two crossings of adiabatic criteria with the logarithmic slope are highlighted by arrows.
    They show large ranges of \gls{itr} for which the adiabatic criteria are below the logarithmic slope: between $\sim0.15$ and $\sim0.32$ for the LP$_{01}$-LP$_{11\mathrm{a}}$ pair, and above $\sim0.38$ for the LP$_{11\mathrm{a}}$-LP$_{11\mathrm{b}}$ pair.
    Those are the three guided modes in the few-mode section, and they are all coupled together via LP$_{11\mathrm{a}}$.
    Hence, the expected behavior is a split of power between the three modes, independent of the illuminated port.
    In addition, the model predicts a quasi-lossless device when fabricated without defects because there is no crossing between any guided and unguided modes. This explanation is provided in more detail in Supplement 1.
    This is witnessed by the fact that the adiabatic criteria coupling the guided modes to the unguided modes of the few-mode section are all above the logarithmic slope curve.
    Although SuPyMode calculates a large amount of adiabatic criteria, for the sake of clarity, we only show the lowest of the guided-to-unguided pairs (between LP$_{01}$ and LP$_{21\mathrm{a}}$) in Fig.~\ref{fig:AAA}(a).
    A more detailed (and more cluttered) example of adiabatic criteria curves is provided as an example in the supplementary material (Supplement 1).

    To ascertain the validity of the predictions from the model, we measured the ratio of power in each output, with respect to the power injected at each input, separately.
    This was done using the in-situ characterization setup at the end of the fusion and tapering process but before cleaving.
    The results are shown in Figs.~\ref{fig:AAA}(b--d).
    The power ratio of each branch with respect to the illumination power is color-coded: branch 1 is blue, 2 is red, and 3 is green). Excess loss at all wavelengths is shown as the black line.
    Illumination of branch 1 corresponds to Fig.~\ref{fig:AAA}(b), branch 2 to Fig.~\ref{fig:AAA}(c), and branch 3 to Fig.~\ref{fig:AAA}(d).

    The three graphs show the expected strong coupling between all the ports.
    The ratio of power sent to the two non-illuminated branches is about -5~\unit{dB}, or one-third, as expected.
    The strong oscillations on the power ratio of the illuminated branch are expected and are due to interference between the modes.
    They are highly dependent on the exact length of the tapered section and do not reflect actual coupling values in the final component (which will be cleaved in the middle of the tapered section).
    Only their average trend is relevant.
    That trend still shows some imperfections of the device, with slightly higher coupling to the non-illuminated branches and slightly higher excess loss (black line) at $\sim1400$~\unit{nm}.
    This results in low ratios of power in the illuminating branch (down to less than 10\% in two cases).
    However, the device displays low overall excess loss.

    Finally, the far-field images of Fig.~\ref{fig:AAA}(e) reinforce the fact that the three modes LP$_{01}$, LP$_{11\textrm{a}}$ and LP$_{11\textrm{b}}$ are well mixed at the output. 
    None of the images features well-defined transverse mode structures.

\begin{figure}[t]
    \centering
    \includegraphics[width=\columnwidth]{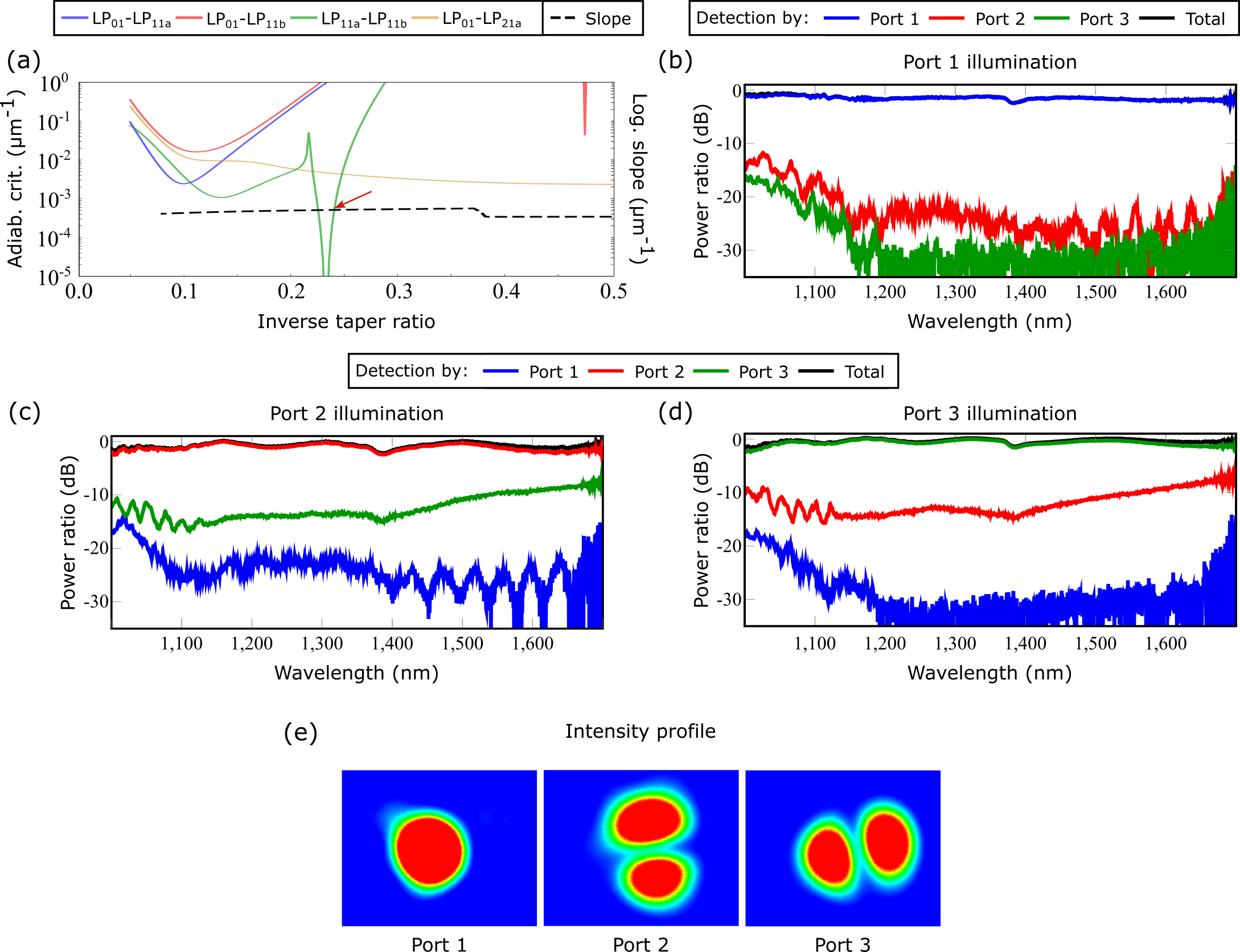}
    \caption{ABC fiber configuration.
    (a) Adiabatic criteria for three different double-clad fibers with 42 (A-branch 1), 32.3 (B-branch 2), and 19.6~\unit{\um} (C-branch 3) first cladding diameter, respectively.
    The solid blue, red, and green lines show the criteria for each pair of guided modes (LP$_{01}$, LP$_{11\mathrm{a}}$, and LP$_{11\mathrm{b}}$) in the few-mode tapered section of the device.
    The lowest adiabatic criterion for coupling between a guided and a non-guided mode (LP$_{01}$ and LP$_{21\mathrm{a}}$ in this instance) is also shown.
    The dashed black curve corresponds to the logarithmic slope for the fabrication sequence.
    (b--d) Power ratio in each branch, with respect to the input power, after fusion and tapering but before cleaving (branch 1 in blue, 2 in red, 3 in green, total in black) when illuminating input 1 (b), input 2 (c), or input 3 (d), respectively.
    (e) Far-field optical profile of the few-mode structure after cleaving and illuminating each port.}
    \label{fig:ABC}
\end{figure}

\subsection{Mode-selective photonic lanterns}\label{sec:ABC}
    \Glspl{mspl} are obtained with components that exhibit no coupling between modes.
    This can be slightly counter-intuitive in the realm of fiber-based components, which often rely on mode-coupling to exhibit interesting behaviors.
    In this case, however, it should be no real surprise as the idea is to segregate the transverse modes of the few-mode section to selectively direct them into three different single-mode fibers.
    This can be readily achieved by sufficiently breaking all the symmetries existing in the previous (AAA) configuration.
    Hence, mode-selectivity is achieved by using three sufficiently different fibers.
    In this study, we chose three different \glspl{dcf} with identical single-mode characteristics (see sub-section~\ref{sec:config}) and varying first-cladding diameters.

    Simulations are displayed in Fig.~\ref{fig:ABC}(a).
    They indicate that there should be no coupling between modes.
    A notable feature, however, seems to suggest otherwise: a crossing of the adiabatic criterion between modes LP$_{11\mathrm{a}}$ and LP$_{11\mathrm{b}}$ at $\sim0.23$ \gls{itr} is highlighted by a red arrow in the graph.
    This is when additional judgment and experience are required in the interpretation.
    This type of very narrow and very abrupt decrease of an adiabatic criterion is usually spurious.
    It is an artifact that exists whenever two propagation constants $\beta_i$ and $\beta_j$ become equal.
    A good rule of thumb is to discard very narrow crossings in the graphs.
    In the end, the simulation suggests that a device fabricated with notable defects should be highly mode-selective and quasi-lossless.

    Figs.~\ref{fig:ABC}~(b--d) show the power transfer for all illumination scenarios where the power remains mainly in the illuminated branch.
    Power transfers to the non-illuminated branches are minimal, below 10\% for most wavelengths.
    Specifically, in Fig.~\ref{fig:ABC}~(b), modal isolation between the LP$_{01}$-like mode and the two LP$_{11}$-like modes is shown to be above 20~\unit{dB} for the range of 1200--1600~\unit{\nano\meter}, and above 12~\unit{\decibel} throughout.
    In Figs.~\ref{fig:ABC}~(c--d), modal isolation between the two LP$_{11}$ modes is shown to be above 10~\unit{\decibel} from 1000--1600~\unit{\nano\meter}.
    In general, isolation is lower, and excess loss is higher in the measured wavelength range.
    The effective operational range of these components remains very large, at around 500~\unit{\nano\meter}.
    In addition, the fabricated device proves to be with low excess loss.

    In contrast to the case of the conventional \gls{pl}, the far-field images of Fig.~\ref{fig:ABC}~(e) display the well-delineated shapes of the expected transverse modes that look like the LP$_{01}$, LP$_{11\mathrm{a}}$, and LP$_{11\mathrm{b}}$.

\begin{figure}[t]
    \centering
    \includegraphics[width=\columnwidth]{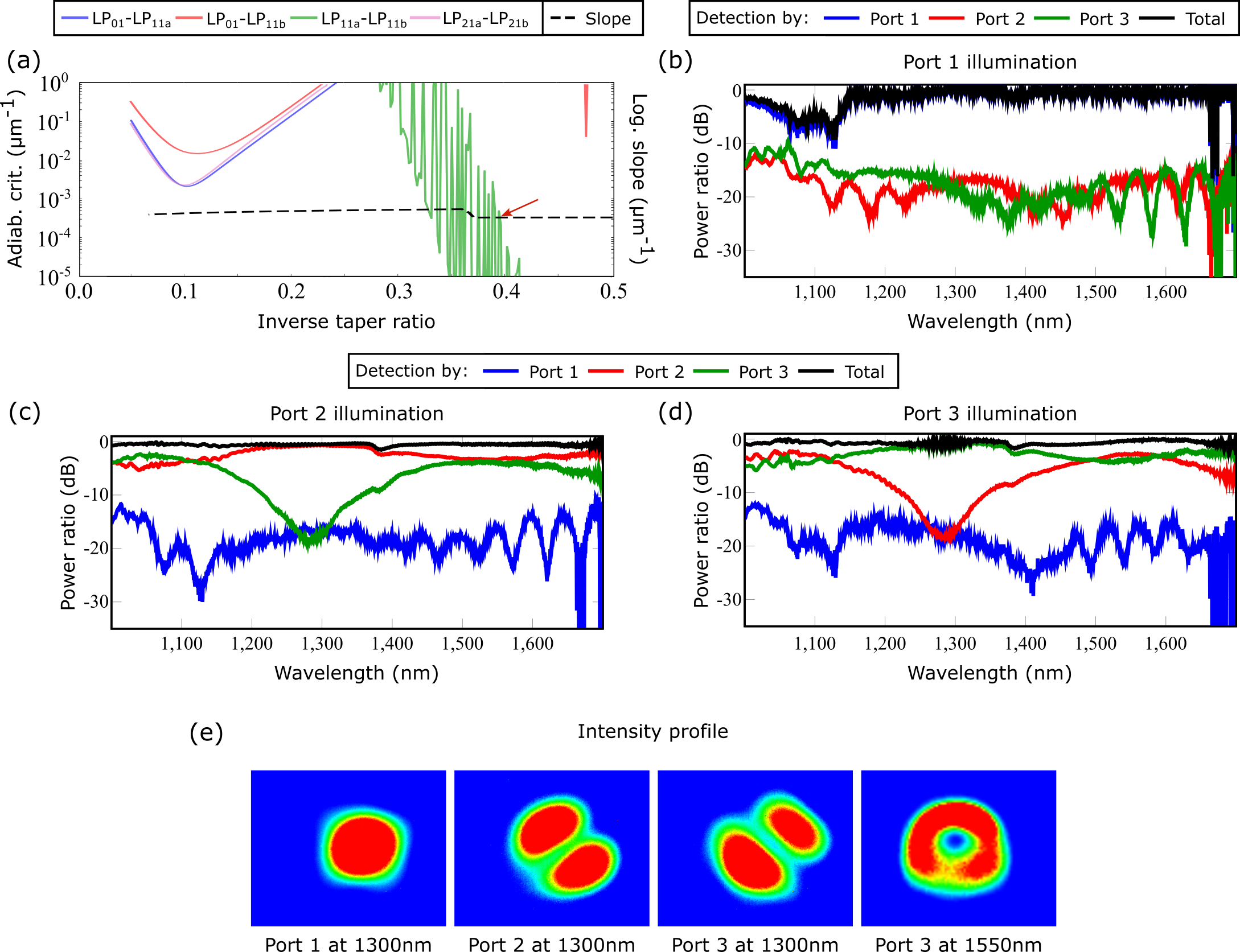}
    \caption{ABB fiber configuration.
    (a) Adiabatic criteria for two different types of double-clad fibers with 42 (A-branch 1) and 32.3 (B-branches 2 and 3) first cladding diameter, respectively.
    The solid blue, red, and green lines show the criteria for each pair of guided modes (LP$_{01}$, LP$_{11\mathrm{a}}$, and LP$_{11\mathrm{b}}$) in the few-mode tapered section of the device.
    The dashed black curve corresponds to the logarithmic slope for the fabrication sequence.
    (b--d) Power ratio in each branch, with respect to the input power, after fusion and tapering but before cleaving (branch 1 in blue, 2 in red, 3 in green, total in black) when illuminating input 1 (b), input 2 (c), or input 3 (d), respectively.
    (e) Far-field optical profile of the few-mode structure after cleaving and illuminating each port. The first three images were obtained at 1300~\unit{nm}. The last one is of port 3 at 1550~\unit{nm}. It shows mixing between the two LP$_{11}$-like modes.}
    \label{fig:ABB}
\end{figure}

\begin{figure}[t]
    \centering
    \includegraphics[width=\columnwidth]{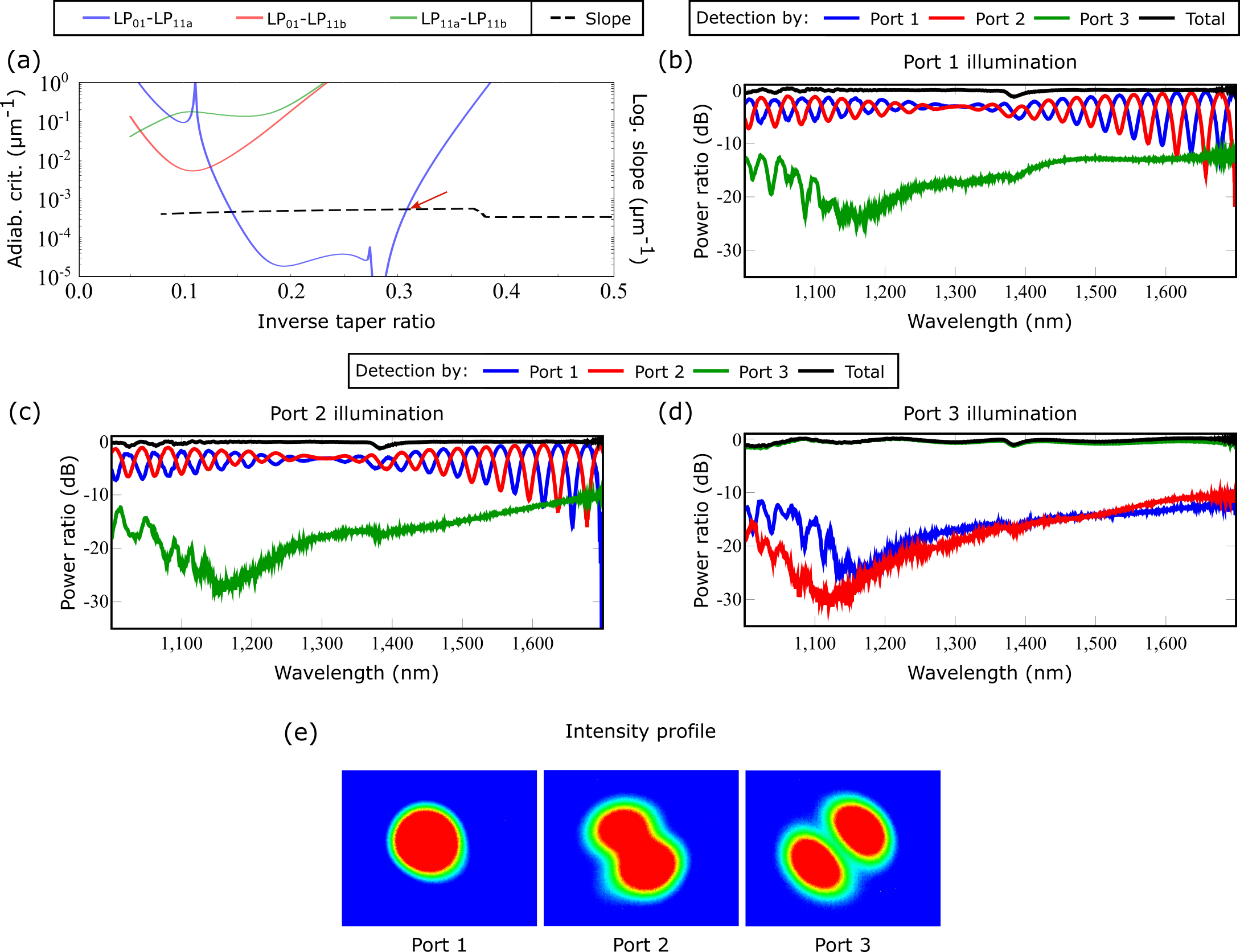}
    \caption{AAB fiber configuration.
    (a) Adiabatic criteria for two different types of double-clad fibers with 42 (A-branches 1 and 2) and 32.3 (B-branch 3) first cladding diameter, respectively.
    The solid blue, red, and green lines show the criteria for each pair of guided modes (LP$_{01}$, LP$_{11\mathrm{a}}$, and LP$_{11\mathrm{b}}$) in the few-mode tapered section of the device.
    The dashed black curve corresponds to the logarithmic slope for the fabrication sequence.
    (b--d) Power ratio in each branch, with respect to the input power, after fusion and tapering but before cleaving (branch 1 in blue, 2 in red, 3 in green, total in black) when illuminating input 1 (b), input 2 (c), or input 3 (d), respectively.
    (e) Far-field optical profile of the few-mode structure after cleaving and illuminating each port.}
    \label{fig:AAB}
\end{figure}

\subsection{Hybrid photonic lanterns}\label{sec:hybrid}
Even with only three fibers in the bundle, devices can be made that will display behaviors between those of conventional and mode-selective components.
An obvious way to achieve such behavior with $3\times1$ \glspl{pl} is to use two different types of fibers.
Here, we introduce two types of hybrid devices, which we call ABB and AAB, where A represents the fiber with the largest first cladding diameter.
Interestingly, the two cases display rather different behaviors.
A device of type ABB will segregate the LP$_{01}$-like mode from the twin LP$_{11}$-like modes, whereas a device of type AAB will hybridize the LP$_{01}$-like mode with one of the LP$_{11}$-like modes, while the other one will be segregated.
Hence, already with 3 fibers, hybrid \glspl{pl} components display a variety that can only grow with the number of fibers and can, in principle, be used to tailor devices to desired applications.

Fig.~\ref{fig:ABB}(a) shows a range of \glspl{itr} above $\sim0.38$ for which the adiabatic criterion between the twin LP$_{11}$ modes is lower than the logarithmic slope curve.
Hence, we can expect a hybridization of those modes and the segregation of the fundamental LP$_{01}$-like mode.
The rapidly oscillating nature of the green adiabatic criterion curve reflects numerical instabilities caused by the degeneracy between the LP$_{11}$ modes.
Only the trend is helpful here.
In contrast, the model does not predict any coupling between the fundamental mode LP$_{01}$ and the twin LP$_{11}$ modes.
Nor does it predict any coupling between guided and unguided modes, which suggests that the particular fabrication sequence, absent defects, can produce a quasi-lossless device.

The experimental results are shown In Fig.~\ref{fig:ABB}(b-e).
In Fig.~\ref{fig:ABB}~(b), illumination through the first port, corresponding to the lone type A fiber and in principle to the fundamental LP$_{01}$-like mode, shows high isolation of that mode from the other two guided modes.
This is close to the scenario of Fig.~\ref{fig:ABC}~(b), although in the case of this specific component, higher losses are observed below 1150~\unit{nm}.
In contrast, much higher power transfers exist between the LP$_{11}$ branches, as shown in Figs.~\ref{fig:ABB}~(c,d).
A slow interferometric beating can erroneously lead to the impression that the coupling is highly dependent on the wavelength.
This is not the case: coupling remains high throughout, and only the exact length of the tapered section dictates, through interference, how much power is found in each branch.

The first three images in Fig.~\ref{fig:ABB}(e) actually show strongly segregated LP$_{11}$-like modes, akin to the mode-selective case shown in Fig.~\ref{fig:ABC}(e).
This is because the wavelength of the narrow-band illumination laser used for this characterization step, 1300~\unit{nm}, is very close to a minimum of the hybridization between the LP$_{11}$-like modes.
For this component, we also took a picture with illumination at 1550~\unit{nm} (last image in Fig.~\ref{fig:ABB}(e)), which shows hybridization of the LP$_{11}$-like modes. 
Interestingly, the interference pattern strongly depends on the polarization state of the illumination laser, a feature that we intend to explore fully in the near future.

A configuration with two A-type fibers (42~\unit{\um} first cladding diameter) and one B-type (32.3~\unit{\um} first cladding diameter), or configuration AAB, is not an intuitively straightforward, as the fundamental mode will be hybridized with one of the LP$_{11}$ modes, leaving the other segregated.
In other words, the break in the physical symmetry does not correspond to the natural symmetry breaking of the mode structure.
The results, shown in Fig.~\ref{fig:AAB} are nonetheless enlightening and we include them here for completeness.

As shown in Fig.~\ref{fig:AAB}(a) there is a large region of \glspl{itr}, between $\sim0.15$ and $\sim0.3$, where the adiabatic criterion of the LP$_{01}$-LP$_{11\text{a}}$ pair falls below the logarithmic slope curve.
We can thus expect power transfers between the two modes while LP$_{11\text{b}}$ remains segregated.

Figs.~\ref{fig:AAB}(b,c) indeed show strong coupling between LP$_{01}$ and LP$_{11\mathrm{a}}$, while LP$_{11\mathrm{b}}$ is isolated.
The rapid interferometric oscillations observed in these figures are highly dependent on the actual length of the tapered section, and only the average value is relevant.
In contrast, illumination of the lone B-type fiber shows modal isolation greater than 10~\unit{dB} throughout.

The optical profiles in the far field, shown in Fig.~\ref{fig:AAB}(e), are interesting.
In particular, the second profile is clearly a superposition of the LP$_{01}$ mode with one of the LP$_{11}$ modes, while the last profile is purely LP$_{11}$-like.
It is unclear if this feature could easily be exploited in an experiment.
    
\section{Discussion}\label{sec:Discussion}
\Gls{cmt} gives rise to very useful modeling tools for the fabrication of complex devices such as \glspl{pl}, as highlighted by the strong agreement between theory and experiment highlighted in this study.
Indeed, we showed that a comparison between adiabatic criteria and logarithmic slopes predicts the behaviors of various types of \glspl{pl}. As with any modeling tool, however, informed judgment is required to interpret results adequately. For instance, the sudden but very brief plunge of an adiabatic criterion can mislead into thinking that coupling will occur when actual modes remain segregated. Such a spurious signal which occurs only when a criterion lies below the logarithmic slope over a large range of \glspl{itr} must be distinguished from actual coupling.

Modeling tools also present limitations, specifically regarding large wavelength bandwidths. Here, each curve plotting the adiabatic criteria vs logarithmic slope is calculated for a single wavelength, and a study over the whole range of intended performance is necessary before attempting to fabricate the device. We did not show such an extensive set of curves in the main text, but an example is provided in the supplementary material~(Supplement 1).

Even with simulation tools, the most comprehensive information about a component's performance is obtained by analyzing the full power transfer for each wavelength performance and its optical profile to determine its usefulness in any application. An important consideration when monitoring results is to avoid misinterpreting segregation as an absence of coupling.
When coupling occurs, two specific modes interfere, and the amount of power within each mode at the output depends on the length of the fully tapered region.
The coupling length needs to be taken into account as the tapered region is subsequently cut in half to make two devices from one fabrication sequence, and no amount of modeling can precisely determine the exact outcome. The performance of each device depends on its coupling efficiency, which is influenced by the effective refractive index of each fiber. Since each fiber has a different first cladding, the effective refractive index changes when the fiber is tapered. This change can break symmetry and selectively generate different modes. However, structural deformations and uneven heating profiles may introduce asymmetries or losses in the system. Therefore, while simulations provide an initial understanding, experimental characterization of each device is essential to account for any changes that might affect coupling efficiency, operational wavelength, and optical profile.

On the other hand, while the AAB configuration demonstrates intriguing behavior, it is crucial to note that it could evolve into a \gls{mspl} by increasing the length of the final component. Fig.~\ref{fig:AAB}(a) illustrates that, by creating a component that reaches values on the order of $10^{-5}$, each mode can be individually generated in the few-mode section. Although this configuration may not yield a highly robust component, it does enable the development of a mode-selective component with a reduced number of fibers.

Finally, all \glspl{pl} fabricated for this study were made with synthetic fused silica capillary tubes, which do not provide the best results in terms of excess loss.
Better components can be manufactured using more expensive fluoride capillary tubes.
With the exception of excess loss, fused silica capillaries allow reproducing the main outcomes obtained with more expensive capillary tubes.

\section{Conclusion}\label{sec:Conclusion}

Fabricating any type of \gls{pl} for its implementation requires identifying their coupling performance, operational wavelength, and optical profile of the output. Therefore, the simulations and characterizations are critical for its implementation. Simulation tools allow for the determination of the essential ingredients needed for manufacturing devices with specific characteristics. This is particularly relevant when the modeling tool indicates no intersection between the adiabatic criteria and the logarithmic slope of the taper profile.

The existence of reliable models allows for the analysis of numerous new \glspl{pl} designs without the need for lengthy trial-and-error manufacturing processes. Meanwhile, the fabrication process and characterization provide insights into the coupling performance, operational wavelength, and optical profile. These performance attributes can vary depending on the fiber type, length of the device, and fabrication procedure.

Additionally, ideally, it is possible to categorize \glspl{pl} into three main types: 
1. Conventional \glspl{pl}, which exhibit strong coupling performance between each port, a high dependence on wavelength, and a superposition of guided modes.
2. Mode-selective devices provide high isolation between all modes, a broadband operational wavelength, and the generation of each mode individually.
3. Hybrid components, which combine the behaviors of the first two types—exhibiting high isolation between certain ports while maintaining strong coupling with others in a large bandwidth with a mixture in the profile of the mode.

\begin{backmatter}
\bmsection{Funding}
Caroline Boudoux and St{\'e}phane Virally acknowledge funding from the Mid-Infrared Quantum Technology for Sensing (MIRAQLS) project, supported by the European Union’s Horizon Europe research and innovation program under grant agreement 101070700. Natural Sciences and Engineering Research Council (NSERC) of Canada grants \#RGPIN-2018-06151 (CB).

\bmsection{Acknowledgment}
The authors thank Mika\"el Leduc for his invaluable work on the fabrication and characterization setup.

\bmsection{Disclosures}
RIBD: Castor Optics, inc. (E), CB: Castor Optics, inc. (I), NG: Castor Optics, inc. (I,P).

\bmsection{Data availability}
The data underlying the results presented in this paper is not publicly available at this time but may be obtained from the authors upon reasonable request.

\section{Supplemental Document}

\subsection{Adiabatic criteria and logarithmic slope}
In the context of fiber devices, adiabaticity refers to the absence of coupling of one specific mode with any other mode. This happens as long as the logarithmic slope \(\frac{1}{\bm{I}}\pdv{\bm{I}}{z}\) remains below the adiabatic criteria $\overline{\alpha}_{ij}$ between the specific mode $i$ and all other modes $j$, at every point along the taper
    \begin{equation}
        \frac{1}{\bm{I}\!(z)}\pdv{\bm{I}\!(z)}{z}<\min_{j}\overline{\alpha}_{ij}(z)\quad\forall z.
        \label{eq:adiabaticcriterion}
    \end{equation}
    
If an adiabatic criterion becomes lower than the logarithmic slope for a significant amount of time, amplitude interference between modes will occur, and power will potentially be transferred from one mode to another. Tools such as SuPyMode compute adiabatic criteria for custom structures at various \glspl{itr}. They can, in turn, be compared to the logarithmic slope expected from the fabrication sequence~\cite{deSivry-Houle:24}.

The computation of the logarithmic slope is described in reference~\cite{birks_shape_1992}, where a longitudinal section of length
    \begin{equation}
        L=L_0+\alpha\,\delta z
    \end{equation}
is heated up. Here, $L_0$ is the initial length of the heated section, $\alpha$ is a unitless coefficient (see below), and $\delta z$ is the total elongation of the component (i.e., the difference between the current length of the component and its initial length). The coefficient $\alpha$ ranges from $-1$ to $+1$: a positive value indicates an increase in the length of the heated segment as the fabrication progresses. A negative value means a decrease of that same length throughout the recipe. The value of $\alpha$ determines the longitudinal shape of the component, following
    \begin{equation}
        \bm{I}\!(z)=\left[1+\frac{2\alpha z}{L_0(1-\alpha)}\right]^{-\frac{1}{2\alpha}},
    \end{equation}
such that the choice of fabrication parameters $L_0$ and $\alpha$ is sufficient to determine the logarithmic slope along the component and, thus, the expected adiabaticity of the component. In general, smaller values of $L_0$ and $\alpha$ yield shorter components, which is beneficial for robustness at the cost of increasing the logarithmic slope and, thus, the potential of coupling between modes.
    
For a \gls{pl} with a final few-mode structure of $\sim10~\unit{\um}$ of diameter, $L_0$ is limited to the range between the size of the heating element (typically around 1~\unit{mm}) and the length at which the component becomes too fragile for practical purposes (typically 10~\unit{mm}).

Furthermore, heuristics dictate that the manufacturability of fused components is limited to adiabatic criteria greater than $10^{-4}$.

Fig.~\ref{SM1} presents the design of a three-mode \gls{mspl} using single-mode fibers. The simulation was performed using SuPyMode~\cite{martin_poinsinet_de_sivry_houle_2023_10215492,deSivry-Houle:24}. Fig.~\ref{SM1}~(a) shows the cross-section of the fiber bundle making up the 3-mode \gls{mspl} after an initial fusion with a degree of fusion of 0.8---the degree of fusion varying between 0 (no fusion) and 1 (fully-fusioned structure).
The core diameters of the single-mode fibers are 8, 10, and $12~\unit{\um}$ for fibers labeled A', B', and C', respectively, with refractive indices equivalent to those of the typical SMF-28 (ITU-T 6.657.A1, Corning, NY, USA) for the core and cladding in all cases. Fig.~\ref{SM1}~(b, top row) shows the field amplitudes of the transverse modes at the single-mode inputs. Fig.~\ref{SM1}~(b, bottom row) shows the corresponding outputs for an \gls{itr}=0.05, i.e., at the end of the tapering process. Note that the scale changes between the top and bottom row of Fig.~\ref{SM1}~(b). At the top, the full-scale is the structure before tapering, while the scale at the bottom is that of the minimum cross-section (waist) after tapering. The first three columns show the three desired modes obtained from selective LP01 illumination.
The last three columns show that other higher-order modes (e.g., LP$_{21}$ and LP$_{02}$) can be obtained by illuminating with non-fundamental modes (LP$_{11}$). In reality, this cannot happen as those modes are unguided in the structure. Instead, coupling to those modes creates loss channels and contributes to excess loss. For the structure simulated in Fig.~\ref{SM1}, the only guided modes of interest are the fundamental mode LP$_{01}$ and the first two higher order modes, LP$_{11_a}$ and LP$_{11_b}$.

Fig.~\ref{SM1}~(c) displays adiabatic criteria computed by \gls{cmt} (in solid lines) and the logarithmic slopes (dashed lines) as a function of \glspl{itr}, computed for different values of $alpha$. In this example, three different taper fabrication recipes are depicted, represented by a null, a positive, and a negative $\alpha$, respectively, each with an initial value of $L_0=10$~\unit{mm}. 
    
    \begin{figure}[htbp]
        \centering
        \includegraphics[width=0.9\textwidth]{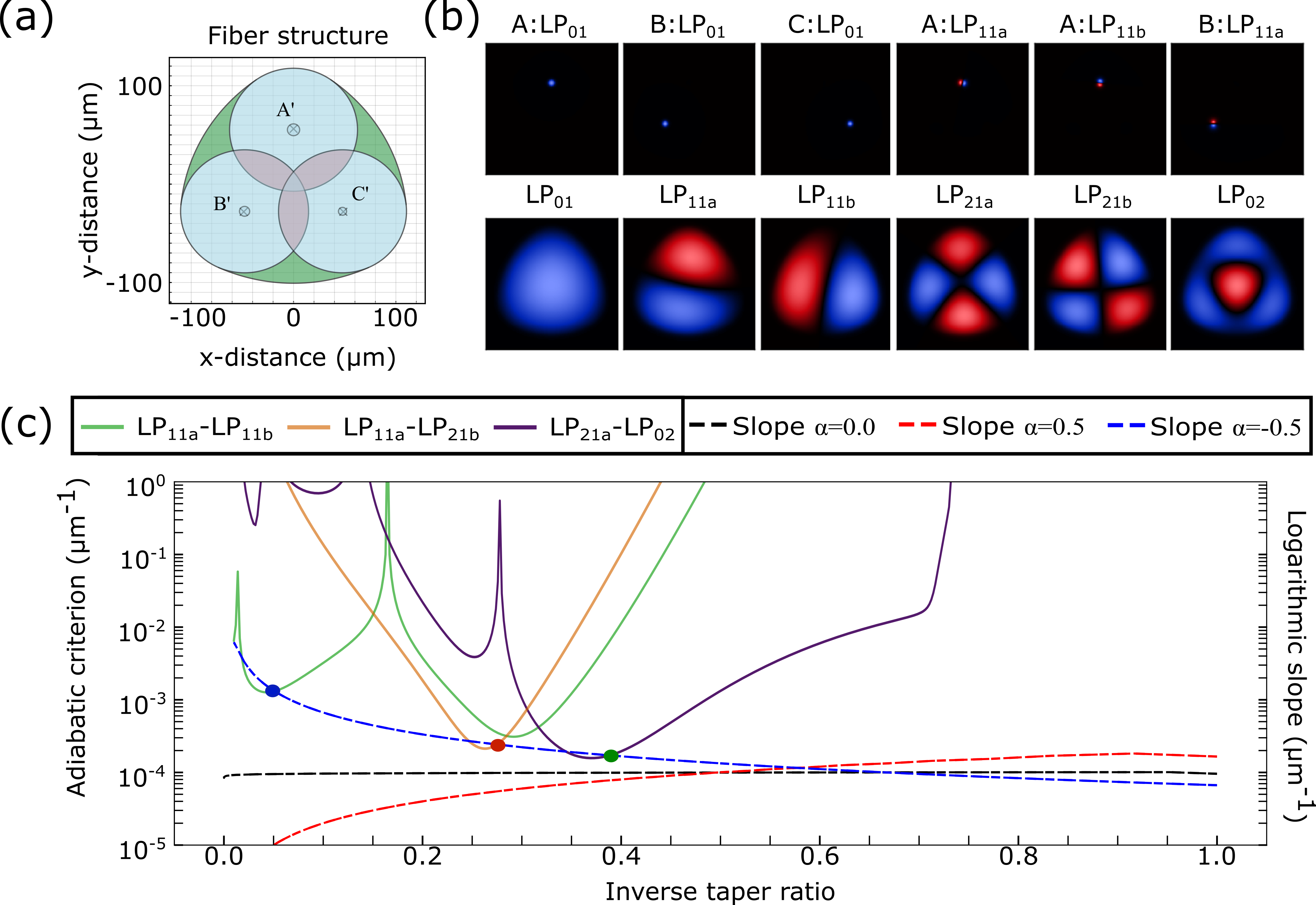}
        \caption{Simulating the design of a 3-mode \gls{mspl} with single-mode fibers.
        (a) Cross-section of the \gls{mspl} after fusion using a fusion coefficient of 0.8.
        (b) Field amplitudes of the transverse modes at the input (first row) and after tapering (\gls{itr}=0.05---second row).
        (c) Adiabatic criteria for each mode pair (solid lines) and logarithmic slopes (dashed lines) for three fabrication sequences using an initial heating length of 10~\unit{mm} and three different $\alpha$ values.
        The red and blue dots show instances of crossing between $\alpha$ and mode mode-coupling curves.}
        \label{SM1}
    \end{figure}

For practical purposes, only the three most relevant curves are plotted into Fig.~\ref{SM1}(c) to highlight the possible responses of the device. In contrast, Fig.~\ref{SM2} presents the typical ``jungle'' of adiabatic criteria produces by the model. Ideally, all modes that interact with the guided modes should be included in the plot. However, adiabatic criteria between higher-order modes tend to be larger, and a cutoff of about 20 pairs is usually implemented in the code to avoid computing too many irrelevant criteria.

\begin{figure}[htbp]
    \centering
    \includegraphics[width=0.9\textwidth]{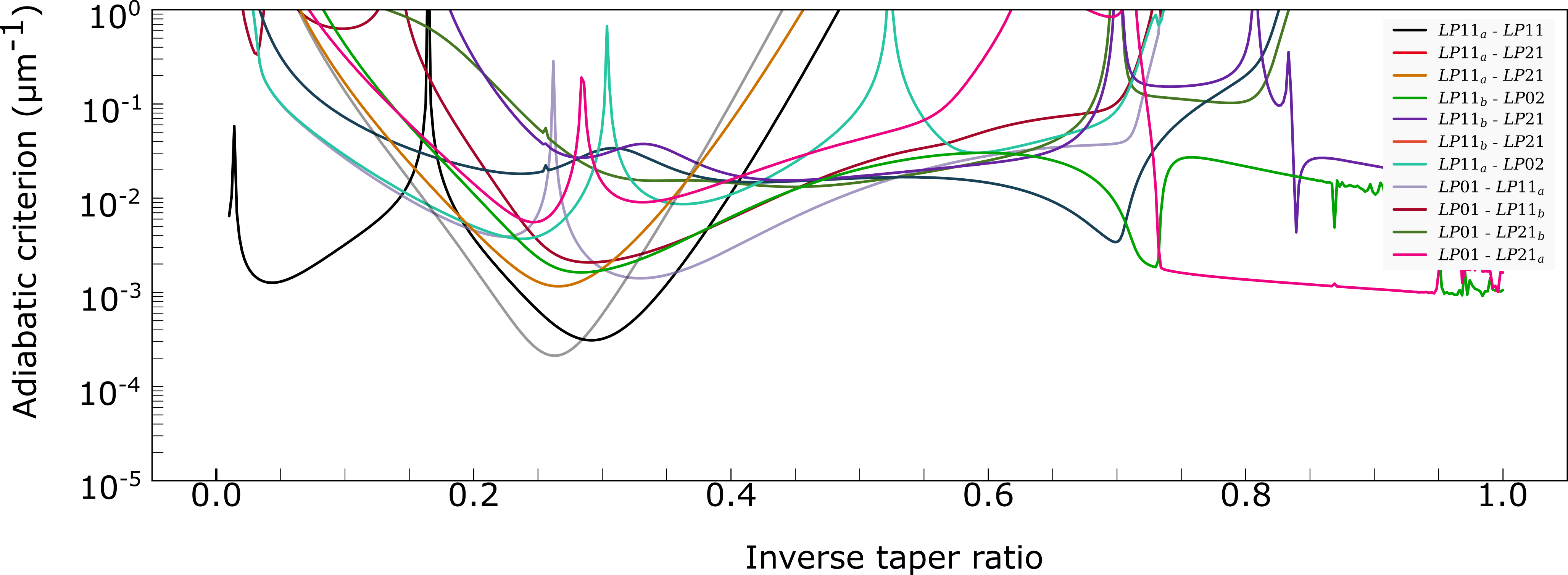}
    \caption{Adibatic criteria for each mode pair of a 3-mode \gls{mspl} with single-mode fibers.}
    \label{SM2}
\end{figure}

Three main crossing types are shown in Fig.~\ref{SM1}(c). They all occur with the blue dashed line, which represents the design with the steepest slopes.
The blue dot corresponds to a crossing between two guided modes, leading to power transferred between the modes. This is intended in some components, such as conventional \glspl{pl} or standard couplers.
The red dot illustrates a crossing between a guided and an unguided modes. These are always to be avoided, as they result in excess loss.
The green dot indicates coupling between two unguided modes, which does not affect the structure at all as these modes do not propagate.
In components such as \glspl{mspl}, all types of crossings are avoided, leading to full modal isolation.
    
\subsection{Further analyses}
To fabricate any component, it is crucial to run simulations at various wavelengths and with different fusion degrees. This is important in particular when the exact degree of fusion cannot be determined in advance, and when the component is intended to be broadband.

Figure \ref{SM3} shows four distinct columns. The first column represents the degree of fusion for a 3-by-1 structure of each row with the fiber configuration ABC, as described in the paper. The second, third, and fourth columns illustrate the adiabatic criteria for the first four pairs of modes at three wavelengths: 1550~\unit{nm}, 1300~\unit{nm}, and 1100~\unit{nm}. This example illustrates an \gls{mspl}, which is the most stable structure. It is intended as a didactical shortcut by simplifying the analysis while still identifying changes in behavior that can arise during the fabrication process. We observe, for instance, that the adiabatic criteria tend to increase with the degree of fusion. Interestingly, at reduced degrees of fusion, coupling starts to occur at smaller ITRs. Finally, as wavelength increases, the areas of potential coupling tend to compactify. It is thus expected that for the same recipe, relatively less coupling will occur at higher wavelengths.
\begin{figure}[htbp]
        \centering
        \includegraphics[width=\textwidth]{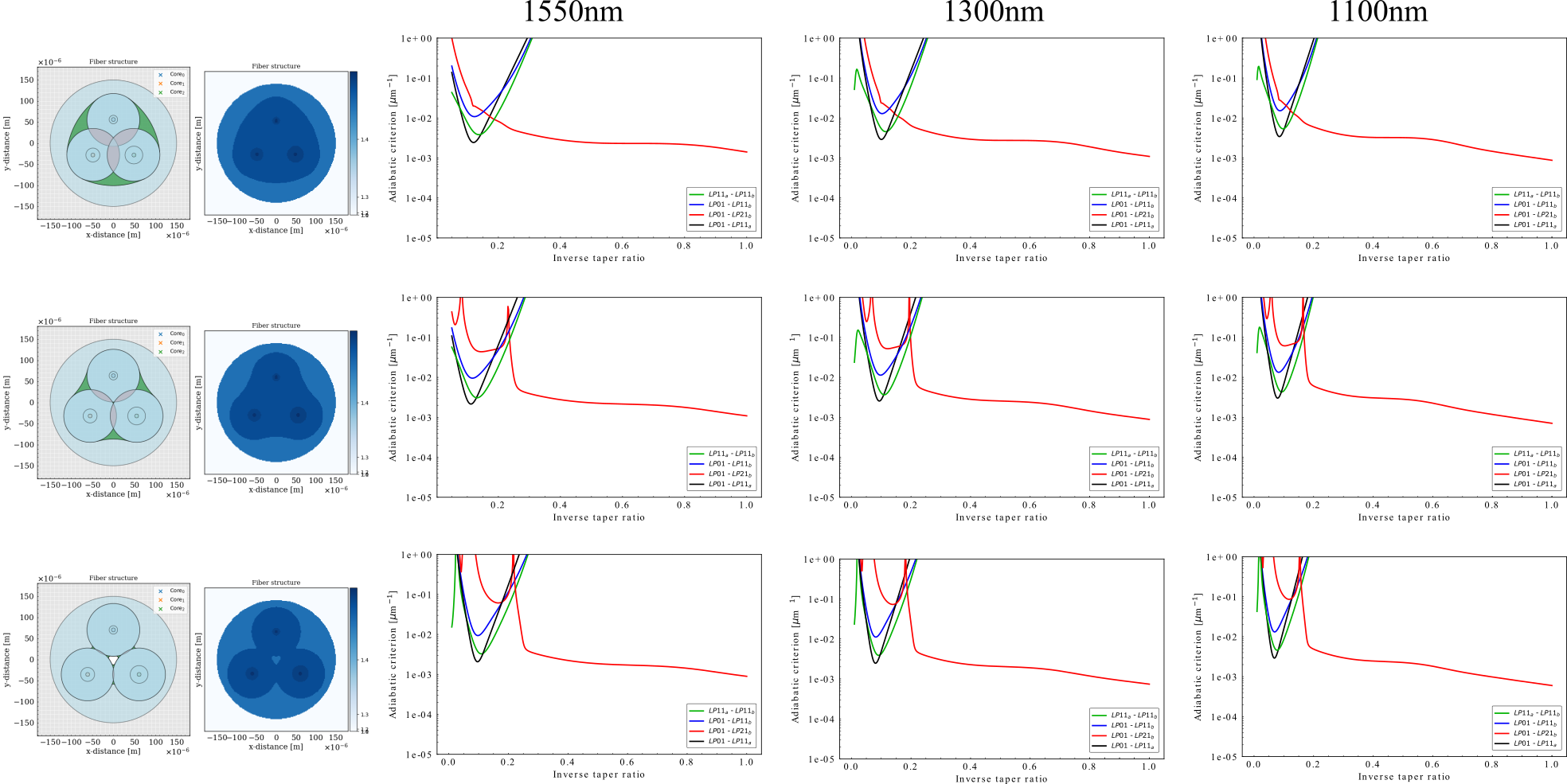}
        \caption{Analysis of the adiabatic criteria at different degrees of fusion and wavelengths. The first column details the degree of fusion corresponding to each row, using the ABC fiber configuration described in the paper. The second, third, and fourth columns display the adiabatic criteria for the first four pairs of modes across three wavelengths: 1550~\unit{nm}, 1300~\unit{nm}, and 1100~\unit{nm}.}
        \label{SM3}
    \end{figure}

Figure \ref{SM4} shows the adiabatic criteria for the first four pairs of modes at three different wavelengths (merging the plots of the three degrees of fusion).

In contrast, Figure \ref{SM5} shows the adiabatic criterion at three distinct degrees of fusion while combining the plots for the three wavelengths.

\begin{figure}[htbp]
    \centering
    \includegraphics[width=\textwidth]{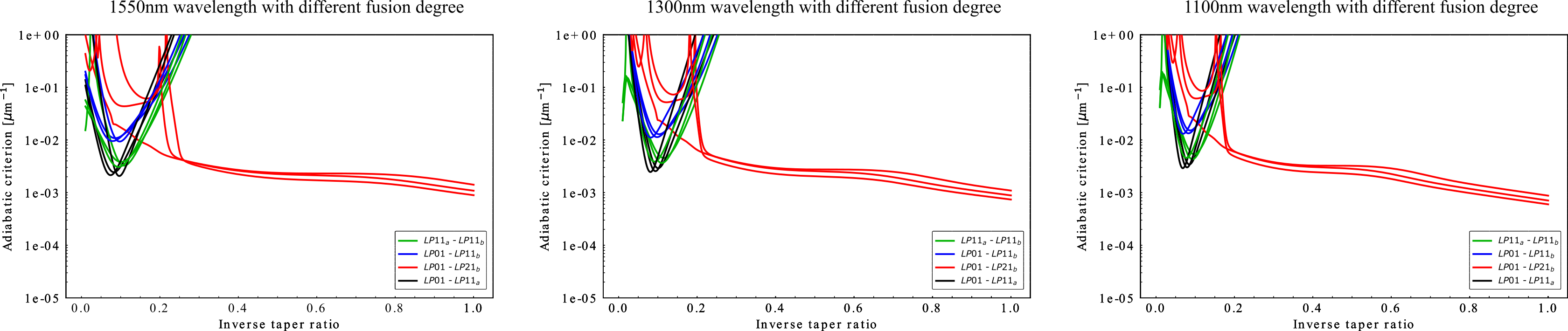}
    \caption{Adibatic criteria of an \gls{mspl} at 1550~\unit{nm}, 1300~\unit{nm}, and 1100~\unit{nm} for the initial four pairs of modes used in combining all the fusion degrees.}
    \label{SM4}
\end{figure}

\begin{figure}[htbp]
    \centering
    \includegraphics[width=\textwidth]{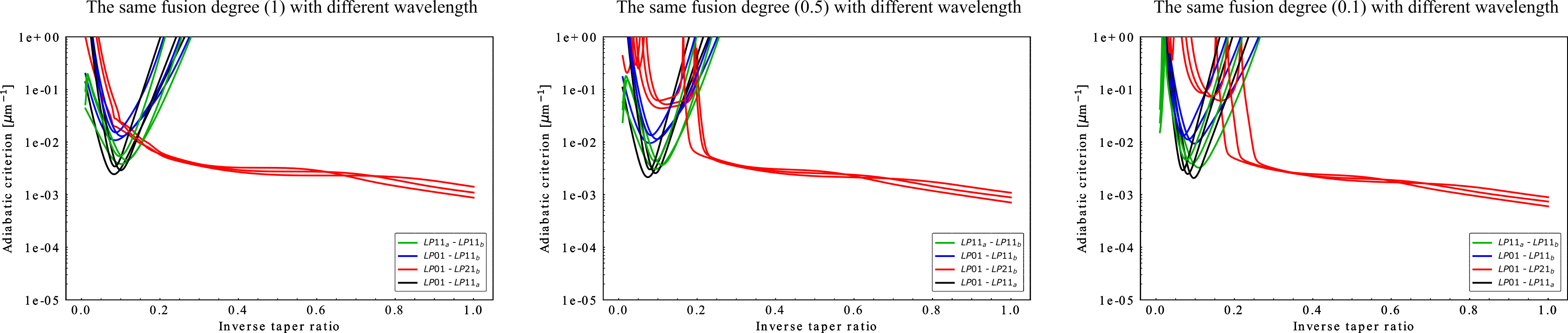}
    \caption{Adibatic criteria for the initial four pairs of modes of an \gls{mspl} at three different fusion degrees, combining the three operational wavelengths: 1550~\unit{nm}, 1300~\unit{nm}, and 1100~\unit{nm}}
    \label{SM5}
\end{figure}

Figure \ref{SM6} shows the adiabatic criteria for all degrees of fusion and all wavelengths. This enables to consolidate our view of all potential variations of behavior during fabrication. A good fabrication sequence for an \gls{mspl} would steer the logarithmic slope away from all the adiabatic criteria curves in this plot.

\begin{figure}[htbp]
    \centering
    \includegraphics[width=0.5\textwidth]{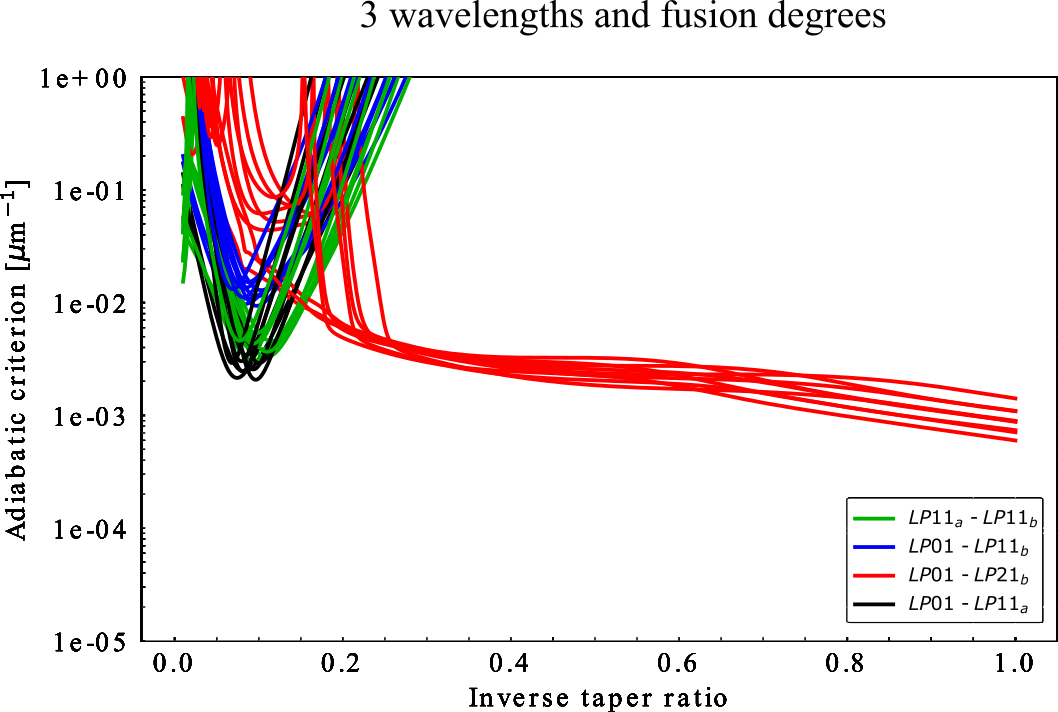}
    \caption{Adibatic criteria for each mode pair of a 3-mode \gls{mspl} at three fusion degrees and wavelengths.}
    \label{SM6}
\end{figure}

\end{backmatter}

\bibliography{bibliography}

\end{document}